\newcommand{\version}{September 7, 2017}
\documentclass[letterpaper,oneside,11pt,pdftex]{article}
\pdfoutput=1
\usepackage[bindingoffset=0.0cm,textheight=22.6cm,hdivide={*,16cm,*}, vdivide={*,22.5cm,*}]{geometry}
\usepackage{amsmath,amsfonts,amssymb}
\usepackage{mathtools}
\usepackage[margin=1.5cm,font=small]{caption}
\usepackage[T1]{fontenc}
\usepackage[utf8]{inputenc}
\usepackage{lmodern}
 \usepackage[numbers,square,comma,sort&compress]{natbib}
 \newcommand{\printbibliography}{\bibliographystyle{utphys-custom} \bibliography{dislocations}}%

\usepackage[pdftex,hyperref,svgnames]{xcolor}
\usepackage[pdftex,bookmarksnumbered=true,breaklinks=true,%
colorlinks=true,linktocpage=true,linkcolor=MediumBlue,citecolor=ForestGreen,urlcolor=DarkRed]{hyperref}

\makeatletter
\AtBeginDocument{
  \hypersetup{
    pdfsubject = {\Abstract},
    pdfkeywords = {\keywords},
    pdftitle = {\@title},
    pdfauthor = {\@author}
  }
}
\makeatother

\makeatletter

 \@addtoreset{equation}{section}
 \makeatother

\renewcommand{\d}{\delta}

\renewcommand{\l}{\lambda}
\newcommand{\m}{\mu}
\newcommand{\n}{\nu}

\newcommand{\s}{\sigma}

 \newcommand{\cH}{\mathcal{H}}

\newcommand{\ml}{l}
\newcommand{\mm}{m}
\newcommand{\mn}{n}

\newcommand{\pa}{\partial}
\newcommand{\inv}[1]{\frac{1}{#1}}

\newcommand{\nn}{\nonumber}
\newcommand{\eqnref}[1]{Eq. \eqref{#1}}

\newcommand{\diff}[2]{\frac{\pa #1}{\pa #2}}

\newcommand{\txt}[1]{\textrm{#1}}

\newcommand{\coleq}{\vcentcolon=}

\title{\texorpdfstring{\begin{flushright}
        {\small LA-UR-17-24288}
       \end{flushright}\vspace{2em}}{}%
       Averaging of elastic constants for polycrystals}

\author{Daniel N. Blaschke}

\date{\version}

\newcommand{\Abstract}{%
Many materials of interest are polycrystals, i.e. aggregates of single crystals.
Randomly distributed orientations of single crystals lead to macroscopically isotropic properties.
Here, we briefly review strategies of calculating effective isotropic second and third order elastic constants from the single crystal ones.
Our main emphasize is on single crystals of cubic symmetry.
Especially the averaging of third order elastic constants has not been particularly successful in the past, and discrepancies have often been attributed to texturing of the polycrystal as well as to uncertainties in the measurement of elastic constants of both poly and single crystals.
While this may well be true, we point out here also shortcomings in the theoretical averaging framework.
}

\newcommand{\keywords}{elastic constants, polycrystals}

\begin{document}

\maketitle

\thispagestyle{empty}
\begin{center}
\vspace{-0.3cm}
Los Alamos National Laboratory,\\
Los Alamos, NM, 87545, USA
\\[0.5cm]
\ttfamily{E-mail: dblaschke@lanl.gov}
\end{center}

%
\begin{abstract}
\Abstract
\end{abstract}


\section{Definition of elastic constants}
\label{sec:elasticconstants}

Assuming no internal torques are present in a crystal, it has been argued that a crystal potential (being rotation and translation invariant) depends only on the Murnaghan strain tensor which is symmetric in its indices (see e.g. \cite[p.~74]{Wallace:1972}, \cite[p.~32]{Hirth:1982},
\cite{Huntington:1958}).
Taylor expanding such a crystal potential in the continuum approximation yields elastic constants with complete Voigt symmetry as its expansion coefficients.
To third order in the strains we get
\begin{align}
 \Phi&=\Phi_0+C_{ij}\eta_{ij}+\inv{2}C_{ijkl}\eta_{ij}\eta_{kl}+\inv{3!}C_{ijklmn}\eta_{ij}\eta_{kl}\eta_{mn}+\ldots
 \label{eq:potential}
\end{align}
where
\begin{align}
 \eta_{ij}=\inv2\left(u_{i,j}+u_{j,i}+u_{k,i}u_{k,j}\right)
\end{align}
is the finite strain tensor of Murnaghan~\cite{Murnaghan:1937} (also known as Lagrangian strain or the Green-Saint-Venant strain tensor~\cite[e.g.][p.~15]{Lubliner:2008}), summation over repeated indices is implied, and $u_{i,j}=\pa_j u_i$ denote the gradients of the continuous displacement field $u_i$.
The linear order term is subsequently eliminated by the equilibrium condition at zero strain (for simplicity we assume here zero initial stress),
\begin{align}
 \diff{\Phi}{\eta_{ij}}\Big|_{\eta=0}=C_{ij}=0
 \,.
\end{align}
The elastic constants $C$ are clearly symmetric in all index pairs and also invariant under exchange of any index pair.
The stress tensor is then derived by
\begin{align}
 \s_{ij}&=\diff{\Phi}{\eta_{ij}}=C_{ijkl}\eta_{kl}+\inv{2}C_{ijklmn}\eta_{kl}\eta_{mn}+\ldots
 \label{eq:stress-strain}
\end{align}
and the inverse strain-stress relation is found to be
\begin{align}
 \eta_{ij}&=\diff{\Psi}{\s_{ij}}=S_{ijkl}\s_{kl}+\inv{2}S_{ijklmn}\s_{kl}\s_{mn}+\ldots
 \label{eq:strain-stress}
\end{align}
where $\Psi=\s_{ij}\eta_{ij}-\Phi$ denotes the Legendre transform of the crystal potential.
Thus, the crystal potential (or strain energy) $\Phi(\eta_{ij})$ depends on the strains $\eta_{ij}$ whereas its Legendre transform, the complementary energy $\Psi(\s_{ij})$, depends on the stress tensors, and each of these energies can be subsequently Taylor expanded in their respective arguments.
Presently, we truncate these expansions at next to leading order so that we may treat third order elastic constants in addition to the second order ones.

The coefficients $S$ are known as ``elastic compliances'', and from Eqs.~\eqref{eq:stress-strain}, \eqref{eq:strain-stress} the following relations are straightforwardly derived~\cite{Lubarda:1997,Kube:2016}:
\begin{align}
 C_{ijkl}S_{klmn}&=\inv2\left(\d_{im}\d_{jn}+\d_{in}\d_{jm}\right)\,,\nn\\
 S_{ijklmn}&=-S_{ijpq}C_{pqrsuv}S_{rskl}S_{uvmn}
 \,. \label{eq:CS-relations}
\end{align}
In general, there are 21 second order elastic constants (SOEC) and 56 third order elastic constants (TOEC)~\cite{Wallace:1970}.
These numbers are reduced by the actual crystal symmetry.
For example, if we consider isotropy, the number of elastic constants is reduced to 2 second and 3 third order constants which are most often parametrized in terms of two Lam{\'e} constants $\l$, $\m$ and three Murnaghan constants $\ml$, $\mm$, $\mn$.

Polycrystals consist of randomly distributed grains of single crystals and can be characterized as ``quasi-isotropic''.
Hence, one may try to derive effective isotropic elastic constants for polycrystals by averaging over the elastic constants of the single crystal grains.
Unfortunately, there is no unique way of doing this~\cite{Barsch:1968,Ballabh:1992,Lubarda:1997,Kube:2016}, and none of the averaging procedures in the literature reproduce the measured effective isotropic third order constants of polycrystals very well~\cite{Kube:2016}.
The situation is further complicated by the fact that many polycrystals can become less isotropic under a variety of conditions, such as when stress is applied, because their grains get aligned to some degree, i.e. they become what is known as textured polycrystals, see e.g.~\cite[pp.~1--9]{Kocks:1998} or~\cite{Wenk:2004,Kube:2016}.

\section{Voigt and Reuss averaging of third order elastic constants}
\label{sec:VR-averaging}

Assuming that all single crystals in a polycrystal are in the same state of strain, Voigt derived the following averaging procedure~\cite{Voigt:1889}, \cite[pp.~954--964]{Voigt:1928}:
Demanding that the following two invariants of the SOEC do not change by the averaging one gets two equations for the two effective Lam{\'e} constants:
\begin{align}
 I_1&=C_{iijj}\,, &
 I_2&=C_{ijij}\,,
\end{align}
and $I^{\txt{averaged}}_m=I^{\txt{single}}_m$.
In using these rotation invariants, one avoids explicitly integrating over all directions.
The same can be done for the third order constants using the additional three invariants
\begin{align}
 I_3&=C_{iijjkk}\,, &
 I_4&=C_{ijijkk}\,, &
 I_5&=C_{ijkijk}
 \,.
\end{align}
For example, if the single crystals are of cubic type, these five equations become
\begin{align}
 I_1&=3(3\l+2\m)=3(c_{11}+2c_{12})\,,\nn\\
 I_2&=3(\l+4\m)=3(c_{11}+2c_{44})\,,\nn\\
 I_3&=6 (9 \ml+\mn)=3 (c_{111}+6 c_{112}+2 c_{123})\,,\nn\\
 I_4&=18 \ml+30 \mm-3 \mn=3 \left(c_{111} +2 (c_{112} +c_{144} +2 c_{166})\right)\,,\nn\\
 I_5&=\frac32 (4 \ml + 20 \mm + \mn)=3 (c_{111} +6 c_{166} +2 c_{456})
 \,,
\end{align}
where the names of the 3 SOEC and 6 TOEC for cubic crystals are as usual inspired by Voigt notation~\cite{Voigt:1889},
\cite[p.~563]{Voigt:1928}.
Explicit tensorial expressions for $C_{ijkl}$, $C_{ijklmn}$ for cubic and isotropic symmetry are given in the appendix.
One immediately finds
\footnote{
A number of authors studied Voigt averaged TOEC for cubic crystals, see e.g.~\cite{Barsch:1968,Chang:1968,Hamilton:1968,Juretschke:1968,Kube:2016}.
}
\begin{align}
 \l^{\txt{V}} &= \frac{1}{5} (c_{11}+4 c_{12}-2 c_{44})\,, \nn\\
 \m^{\txt{V}} &= \frac{1}{5} (c_{11}-c_{12}+3 c_{44})\,, \nn\\
  \ml^{\txt{V}} &= \frac{1}{70} \left(3 c_{111}+26 c_{112}+6 c_{123}+8 (c_{144}- c_{166}- c_{456})\right)\,,\nn\\
  \mm^{\txt{V}} &= \frac{1}{35} (3 c_{111}-2 c_{112}-c_{123}+c_{144}+20 c_{166}+6 c_{456})\,,\nn\\
  \mn^{\txt{V}} &= \frac{4}{35} \left(c_{111}-3 c_{112}+2 c_{123}+9(-c_{144}+ c_{166}+ c_{456})\right)
  \,,
\end{align}
where the superscript refers to Voigt averaging.

If on the other hand, one assumes that the single crystals are in the same state of stress, one may use the five invariants
\begin{align}
 J_1&=S_{iijj}\,, &
 J_2&=S_{ijij}\,, \nn\\
 J_3&=S_{iijjkk}\,, &
 J_4&=S_{ijijkk}\,, &
 J_5&=S_{ijkijk}
 \,,
\end{align}
of the compliances to derive the average isotropic constants demanding $J^{\txt{averaged}}_m=J^{\txt{single}}_m$.
This averaging procedure was first suggested and carried out for SOEC by Reuss~\cite{Reuss:1929}.
The derivation is equally straightforward, though slightly more cumbersome due to the use of relations \eqref{eq:CS-relations}, and yields the Reuss averages
\begin{align}
 \l^{\txt{R}} &= \frac{c_{11}^2+c_{11} c_{12}-2 c_{11} c_{44}-2 c_{12}^2+6 c_{12} c_{44}}{3 (c_{11}- c_{12})+4 c_{44}} \,, \nn\\
 \m^{\txt{R}} &= \frac{5 c_{44} (c_{11}-c_{12})}{3 (c_{11}- c_{12})+4 c_{44}} \,, \nn\\
  \ml^{\txt{R}} &= \frac{c_{111}+6 c_{112}+2 c_{123}}{18} - \frac{n^{\txt{R}}}9 \,,\nn\\
  \mm^{\txt{R}} &= \frac{15 (c_{11}-c_{12})^2 (c_{144}+2 c_{166})+20 c_{44}^2 (c_{111}-c_{123})}{3 (3 (c_{11}- c_{12})+4 c_{44})^2} +\frac{n^{\txt{R}}}{6} \,,\nn\\
  \mn^{\txt{R}} &= \frac{200 c_{44} \Big(4 c_{44}^2 (c_{111}-3 c_{112}+2 c_{123})-9 (c_{11}-c_{12})^2 (c_{144}-c_{166})\Big)+900 c_{456} (c_{11}-c_{12})^3}{7 (3 (c_{11}- c_{12})+4 c_{44})^3}
  \,,
\end{align}
for SOEC and TOEC
\footnote{
A number of authors studied Reuss averaged TOEC for cubic crystals, see e.g.~\cite{Barsch:1968,Chang:1968,Hamilton:1968,Kube:2016}.
}.

At second order, the Voigt and Reuss averages yield bounds on the actual effective Lam{\'e} constants.
Therefore, in taking the mean of both averages some improvement may be achieved~\cite{Hill:1952}.
In particular, one always has $\l^{\txt{R}}>\l^{\txt{V}}$ and $\m^{\txt{V}}>\m^{\txt{R}}$.
This can be easily seen by rewriting $\m^\txt{V}$ and $\m^\txt{R}$ in terms of Zener's anisotropy ratio, $A\coleq 2c_{44}/(c_{11}-c_{12})$, and taking their difference:
\begin{align}
 \m^\txt{V}-\m^\txt{R}&=\frac{c_{44}}5\left(\frac{2}{A}+3\right)-\frac{5 c_{44}}{(3+2A)}
 =\frac{6c_{44}}{5A(3+2A)}\left(1-A\right)^2 \ge 0
 \,,
\end{align}
i.e. since both $c_{44}$ and $A$ are positive,
$\m^\txt{V}-\m^\txt{R}$ is always greater than zero
\footnote{
Tables \ref{tab:values-metals-cubic}, \ref{tab:values-metals-Voigt}, and \ref{tab:values-metals-Reuss} highlight this property for four metals: Al, Cu, Fe for anisotropy ratio $A>1$, and Nb for $A<1$.
In all these cases we find $\m^{\txt{V}}>\m^{\txt{R}}$ and $\l^{\txt{V}}<\l^{\txt{R}}$.
}.
The second relation, $\l^{\txt{R}}>\l^{\txt{V}}$, then immediately follows from $\l=K-2\m/3$ together with $K^{\txt{V}}=K^{\txt{R}}$ (which we show below).

At third order, unfortunately, the two averages (Voigt and Reuss) do not yield bounds on the true average in a polycrystal limiting their usefulness.
In Tables \ref{tab:values-metals-Voigt} and \ref{tab:values-metals-Reuss} we compare Voigt and Reuss averages, respectively, using single crystal data listed in Table~\ref{tab:values-metals-cubic} to experimental values for polycrystals listed in Table~\ref{tab:values-metals-isotropic}.
We have chosen a selection of cubic fcc and bcc crystals with anisotropy factors ranging from $0.5<A<3.27$ due to availability of both single crystal and polycrystal data:
They are aluminum, copper, iron, and niobium.

\begin{table}[h!t!b]
{\renewcommand{\arraystretch}{1.1}
\centering
 \begin{tabular}{l|r|r|r|r}
          & Al\,(fcc) & Cu\,(fcc) & Fe\,(bcc) & Nb\,(bcc) \\\hline
 $c_{11}$[GPa]\!\! & $106.75\pm0.05$ & 166.1 & $226\pm2$ & $246.5\pm0.5$ \\
 $c_{12}$[GPa]\!\! & $60.41\pm0.08$ & 119.9 & $140\pm8$ & $133.3\pm0.7$ \\
 $c_{44}$[GPa]\!\! & $28.34\pm0.04$ & 75.6 & $116\pm1$ & $28.4\pm0.06$ \\
 $c_{111}$[GPa]\!\! & $-1076\pm30$ & $-1271\pm22$ & $-$2720 & $-2564\pm25$ \\
 $c_{112}$[GPa]\!\! & $-315\pm10$ & $-814\pm9$ & $-$608 & $-1140\pm25$ \\
 $c_{123}$[GPa]\!\! & $36\pm15$ & $-50\pm18$ & $-$578 & $-467\pm25$ \\
 $c_{144}$[GPa]\!\! & $-23\pm5$ & $-3\pm9$ & $-$836 & $-343\pm10$ \\
 $c_{166}$[GPa]\!\! & $-340\pm10$ & $-780\pm5$ & $-$530 & $-167.7\pm5$ \\
 $c_{456}$[GPa]\!\! & $-30\pm30$ & $-95\pm87$ & $-$720 & $136.6\pm5$
 \end{tabular}
 \caption{We list the experimental values used in the computation of the averages:
 In particular, SOEC and TOEC at room temperature for Al are taken from~\cite{Thomas:1968},
 those for Cu are taken from~\cite{Hiki:1966},
 those for Fe are taken from~\cite{Leese:1968} and~\cite{Powell:1984},
 and those for Nb are taken from~\cite{Graham:1968}.
 Whenever uncertainties were given in the references, we have printed them in the table above as well.}
 \label{tab:values-metals-cubic}
}
\end{table}

\begin{table}[h!t!b]
{\renewcommand{\arraystretch}{1.1}
\centering
 \begin{tabular}{c|c|c|c|c}
          & Al\,(fcc) & Cu\,(fcc) & Fe\,(bcc) & Nb\,(bcc) \\\hline
 $\lambda$[GPa]\!\! & 58.1 & 105.5 & 115.5 & 144.5 \\
 $\mu$[GPa]\!\! & 26.1 & 48.3 & 81.6 & 37.5 \\\hline
 $\ml$[GPa]\!\! & $-143\pm13$ & $-160\pm70$ & $-170\pm40$ & $-610\pm80$ \\
 $\mm$[GPa]\!\! & $-297\pm6$ & $-620\pm10$ & $-770\pm10$ & $-220\pm30$ \\
 $\mn$[GPa]\!\! & $-345\pm4$ & $-1590\pm20$ & $-1520\pm10$ & $-300\pm20$ 
 \end{tabular}
 \caption{We list the experimental values for polycrystals which we compare our averages to.
 The Lam\'e constants were taken from Refs.~\cite{Lubarda:1997}, \cite[p.~10]{Hertzberg:2012}.
 The Murnaghan constants for Cu and Fe were taken from~\cite{Seeger:1960}, those for Al were taken from Reddy 1976 as reported by Wasserb{\"a}ch in Ref.~\cite{Wasserbaech:1990}, and those for Nb were finally taken from~\cite{Graham:1968}.
 Uncertainties (as given in those references) are listed as well.}
 \label{tab:values-metals-isotropic}
}
\end{table}

\begin{table}[h!t!b]
{\renewcommand{\arraystretch}{1.1}
\centering
 \begin{tabular}{c|c|c|c|c}
          & Al & Cu & Fe & Nb \\\hline
 $\lambda$[GPa]\!\! & 58.3 & 98.9 & 110.8 & 144.6 \\
 $\mu$[GPa]\!\! & 26.3 & 54.6 & 86.8 & 39.7 \\\hline
 $\ml$[GPa]\!\! & -120 & -261 & -345 & -609 \\
 $\mm$[GPa]\!\! & -275 & -523 & -632 & -223 \\
 $\mn$[GPa]\!\! & -364 & -775 & -660 & 312 
 \end{tabular}
 \caption{We list the Voigt averages for polycrystals using the single crystal data presented in Table~\ref{tab:values-metals-cubic}.}
 \label{tab:values-metals-Voigt}
}
\end{table}

\begin{table}[h!t!b]
{\renewcommand{\arraystretch}{1.1}
\centering
 \begin{tabular}{c|c|c|c|c}
          & Al & Cu & Fe & Nb \\\hline
 $\lambda$[GPa]\!\! & 58.5 & 108.9 & 122.6 & 147.4 \\
 $\mu$[GPa]\!\! & 26.0 & 39.6 & 69.1 & 35.5 \\\hline
 $\ml$[GPa]\!\! & -123 & -373 & -312 & -624 \\
 $\mm$[GPa]\!\! & -269 & -287 & -661 & -192 \\
 $\mn$[GPa]\!\! & -342 & 227 & -950 & 448
 \end{tabular}
 \caption{We list the Reuss averages for polycrystals using the single crystal data presented in Table~\ref{tab:values-metals-cubic}.}
 \label{tab:values-metals-Reuss}
}
\end{table}

It has also been known for some time that a particular combination of the second order constants, the bulk modulus $K=\l+2\m/3$, is uniquely defined~\cite{Barsch:1968} in terms of cubic single crystal constants, i.e. $K^{\txt{V}}=K^{\txt{R}}$.
The reason is actually quite simple:
The bulk modulus describes the materials reaction to isotropic pressure at second order, i.e. $\s_{ij}=\s\d_{ij}$ with $\s=-P$.
The symmetry properties of the elastic constants entering the two expansions \eqref{eq:stress-strain}, \eqref{eq:strain-stress} for cubic symmetry immediately yield strains of the same tensorial form: $\eta_{ij}=\eta\d_{ij}$.
Hence, inserting these special cases of strain and stress and taking the trace we get the following simpler set of equations:
\begin{align}
 3\s&=C_{iikk}\eta+C_{iikkmm}\eta^2
 \,,&
 3\eta&=S_{iikk}\s+S_{iikkmm}\s^2
 \,.
\end{align}
We therefore find the relations
\begin{align}
 S_{iikk}&=9/C_{iikk}\,, &
 S_{iikkmm}&=-C_{iikkmm}\left(\frac{3}{C_{iikk}}\right)^3
 \,, \label{eq:newrelations}
\end{align}
which tell us that two of the five invariants used in the Voigt averaging are related to their counterparts in Reuss averaging.
In particular, $I_1$ and $I_3$ are related to $J_1$ and $J_3$, telling us that two linear combinations of the five effective isotropic constants are uniquely defined.
In particular, for isotropic elastic constants $C_{iikk}=I_1=9K$ is proportional to the bulk modulus reproducing the known result above.
What is equally trivial to derive, but seems not to be so well known is the fact that due to the second relation in \eqref{eq:newrelations} above, the combination of TOEC
\begin{align}
 (9\ml^{\txt{V}}+\mn^{\txt{V}})=(9\ml^{\txt{R}}+\mn^{\txt{R}})
 =\frac12(c_{111}+6 c_{112}+2 c_{123})
 \,,\label{eq:thirdorder-unique}
\end{align}
is also uniquely defined~\cite{Barsch:1968}
for cubic single crystals.
Hence, the problem of finding good averages for TOEC for polycrystals whose grains are crystals of cubic symmetry reduces to finding two of the three effective Murnaghan constants.

In Table~\ref{tab:values-metals-invariants} we compare these uniquely defined averages to their experimental counterparts.
We see that the averages for the bulk modulus agree fairly well, while the averages for the combination of TOEC, $(\ml+\mn/9)$, is merely in the ballpark of the corresponding experimental values, and part of the discrepancies can be attributed to experimental uncertainties~\cite{Wasserbaech:1990,Lubarda:1997,Kube:2016}.
The agreement here, however, is significantly better than for some of the other averaged TOEC, see Tables \ref{tab:values-metals-Voigt}, \ref{tab:values-metals-Reuss}, \ref{tab:values-metals-Lubarda}, and \ref{tab:values-metals-myaverages}.

\begin{table}[h!t!b]
{\renewcommand{\arraystretch}{1.1}
\centering
 \begin{tabular}{c|c|c|c|c}
          & Al & Cu & Fe & Nb \\\hline
 $K_\txt{exp.}$[GPa]\!\! & 75.5 & 137.7 & 169.9 & 169.5 \\
 $K_\txt{av.}$[GPa]\!\! & 75.9 & 135.3 & 168.7 & 171.0 \\
 $(\ml+\mn/9)_\txt{exp.}$[GPa]\!\! & -181 & -337 & -339 & -643 \\
 $(\ml+\mn/9)_\txt{av.}$[GPa]\!\! & -161 & -348 & -418 & -574
 \end{tabular}
 \caption{We compare averaged to experimental values for the bulk modulus $K=\l+2\m/3$ as well as for the corresponding third order quantity $\ml+\mn/9$ to experimental values using the single crystal data presented in Table~\ref{tab:values-metals-cubic} and the polycrystal data presented in Table~\ref{tab:values-metals-isotropic}.
 Note that all averaging schemes discussed above lead to the same numbers for the quantities in this table.}
 \label{tab:values-metals-invariants}
}
\end{table}

\section{Improved averaging procedures of elastic constants}
\label{sec:improved averaging}

For SOEC, there exist improved bounds derived by Hashin and Shtrikman, see Ref.~\cite{Hashin:1961a,Hashin:1962a,Hashin:1962b}.
Additionally, a ``self-consistent'' method was developed by Hershey~\cite{Hershey:1954} and Kr{\"o}ner~\cite{Kroener:1958} using an earlier result by Eshelby~\cite{Eshelby:1957} for SOEC (see also Ref.~\cite[pp.~421--439]{Mura:1987} and
\cite[pp.~282--324]{Kocks:1998}).
Lubarda~\cite{Lubarda:1997} later extended this method to what he calls the ``semi-self-consistent'' method for TOEC.
An alternative method based on simulating clusters of crystallites was presented in~\cite{Kiewel:1996}.
See also~\cite{Ballabh:1992,Brenner:2009,Kube:2016} for recent treatments of averaged elastic constants for polycrystals and additional references.

We now revisit Lubardas treatment (which is based on the works of Hershey and Kr{\"o}ner) of polycrystals whose single crystals have cubic symmetry.
The crucial observation is that in departing from isotropy also the strains receive corrections, i.e. $\eta_{ij}=\eta^0_{ij}+\d\eta_{ij}$ in the expansion \eqref{eq:stress-strain} above.
Using the simplifying assumption that the strain in a single crystal of a polycrystalline aggregate is proportional to the applied strain, one finds~\cite{Lubarda:1997}
\begin{align}
\eta^0_{ij}&=\cH_{ijkl}\eta_{kl}\,,\nn\\
\cH_{ijkl}&=\frac12\left(\d_{ik}\d_{jl}+\d_{il}\d_{jk}\right)+h\left(\d_{ij}\d_{kl}+\d_{ik}\d_{jl}+\d_{il}\d_{jk}-5A_{ijkl}\right)\,,\nn\\
h&=\frac{(c_{11}+2c_{12}+6\m)(c_{11}-c_{12}-2\m)}{3\left[8\m^2+9c_{11}\m+(c_{11}-c_{12})(c_{11}+2c_{12})\right]}\,,\nn\\
A_{ijkl}&=a_ia_ja_ka_l+b_ib_jb_kb_l+c_ic_jc_kc_l
\,, \label{eq:corrected-strain}
\end{align}
where $a_i$, $b_i$, $c_i$ are the orthogonal unit vectors along the principal cubic axis.
Note that $h$ is proportional to the anisotropy factor $(c_{11}-c_{12}-2\m)$ of the second oder elastic constants and hence goes to zero in the isotropic limit.
Using the corrected strains above, one defines
\begin{align}
\hat C_{ijkl}&=C_{ijpq}\cH_{pqkl}\,,\nn\\
\hat C_{ijklmn}&=C_{ijpqrs}\cH_{pqkl}\cH_{rsmn}
\,, \label{eq:defHattedCs}
\end{align}
and we now have to equate the invariants of $\hat C$ to their isotropic counterparts along the lines of the previous section.

The second order hatted elastic constant have the form
\begin{align}
 \hat C_{ijkl}&=(1+2h)C_{ijkl}+hC_{ijpp}\d_{kl}-5hC_{ijpq}A_{pqkl}\nn\\
 &=C_{ijkl}+h\Big((c_{11}-c_{12})\d_{ij}\d_{kl}+2c_{44}(\d_{ik}\d_{jl}+\d_{il}\d_{jk})-(3(c_{11}-c_{12})+4c_{44})A_{ijkl}\Big)
\end{align}
leading to the invariants
\begin{align}
 \hat C_{iikk}&=
3(c_{11}+2c_{12})
 \,,\nn\\
 \hat C_{ikik}
 &=3(c_{11}+2c_{44})+6h(2c_{44}+c_{12}-c_{11})
 \,, \label{eq:SOECInvarsLubarda}
\end{align}
where we used
\footnote{
If we choose our coordinates such that $a_i=\hat x_i$, $b_i=\hat y_i$, and $c_i=\hat z_i$, then we immediately see that $A_{pqkk}=a_ia_j+b_ib_j+c_ic_j=\d_{ij}$.
Since $\d_{ij}$ is an invariant tensor, this relation is true even when our coordinates are not aligned with the cubic axes.
}
\begin{align}
 C_{iikk}&=3(c_{11}+2c_{12})\,,
 &
 C_{ijij}&=3(c_{11}+2c_{44})
 \,,\nn\\
 A_{ijkk}&=a_ia_j+b_ib_j+c_ic_j=\d_{ij}
\,,&
\cH_{ijkk}&
=\d_{ij} 
\,.
\end{align}
Notice that $\hat C_{iikk}=C_{iikk}$ since this quantity is proportional to the bulk modulus, i.e. this new averaging scheme will again lead to the same value for the bulk modulus as before (as it should be).
Hence, as before $\l$ is given in terms of the averaged bulk modulus and $\m$, i.e.
\begin{align}
 \l&=K-2\m/3=(c_{11}+2c_{12}-2\m)/3
 \,.
\end{align}
Furthermore, \eqnref{eq:SOECInvarsLubarda} leads to the following cubic equation for the average of $\m$ (first derived by Kr{\"o}ner~\cite{Kroener:1958}, 
see also~\cite{Lubarda:1997}):
\begin{align}
 8 \mu ^3+\mu ^2 (5 c_{11}+4 c_{12})+ \mu  (4 c_{12}-7 c_{11}) c_{44}-c_{44} (c_{11}-c_{12}) (c_{11}+2 c_{12})=0
 \,.
\end{align}
Only one of the three solutions is a positive real for $c_{11}>c_{12}>c_{44}>0$, and hence the present method yields a unique solution for the averaged polycrystalline Lam{\'e} constants.

At third order, Lubarda~\cite{Lubarda:1997} considers the following three invariants to derive his solution for the averaged Murnaghan constants:
$C_{iijjkk}$, $C_{kkijij}$, and $C_{ijkijk}$.
The first of these invariants is independent of $h$ and (as expected) yields once more the same average as \eqnref{eq:thirdorder-unique}.

The other two invariants then yield averages for the remaining two elastic constants.
However, this set of invariants is not unique:
Because of how $\hat C_{ijklmn}$ was defined in \eqnref{eq:defHattedCs} it no longer has Voigt symmetry in the sense that exchanging the first index pair with any other does not lead to the same expression.
Therefore, $C_{kkijij}\neq C_{ijijkk}$ and it makes a difference whether the first or one of the other two index pairs is traced.
The author of Ref.~\cite{Lubarda:1997} chooses to trace the first index pair leading to corrections which are second order in $h$, whereas if one were to trace one of the other index pairs, the corrections would be only linear in $h$ (and half the value of the other case to linear order).

Another point of criticism is that the corrections which are second order in the anisotropy $h$ are kept rather than discarded in Ref.~\cite{Lubarda:1997}:
Since the starting point of the present derivation, \eqnref{eq:corrected-strain}, was based on linear order corrections to the strain, one should consistently discard all higher order terms in $h$.

Similar considerations apply to the compliances approach including linear corrections to the stress field:
Once more the third order hatted compliances loose Voigt symmetry leading to four instead of three invariants, two of them depending on $h^2$ which should be discarded.

Thus, the averaging procedure for TOEC of Ref.~\cite{Lubarda:1997} is really a collection of possible averaging procedures, all leading to different results.
Furthermore, none seem to agree any better with experiments than the Voigt or Reuss averages, i.e. depending on the material and which of the three Murnaghan constants is compared, any one of these averaging procedures may perform ``best''.
Some of the discrepancies have in the past been attributed to texturing~\cite{Kube:2016}.
In fact, the authors of Ref.~\cite{Kube:2016} have even gone so far as to propose to take the simple Hill average (i.e. the mean between Voigt and Reuss averages) since none of the previously proposed averaging schemes have been particularly successful in reproducing TOEC of polycrystals.
On top of these difficulties, the experimental data both for polycrystals and single crystals is wrought with uncertainties for TOEC~\cite{Wasserbaech:1990,Lubarda:1997,Kube:2016}.
Additionally, the single crystal data listed in Table~\ref{tab:values-metals-cubic} were measured using a method developed in Ref.~\cite{Thurston:1964} and do not take into account later improvements of Refs.~\cite{Sekoyan:1975,Sekoyan:1983} (which appeared years after those measurements).
Hence, those single crystal data are likely to contain significant systematic errors~\cite{Skove:1967,Powell:1967,Markenscoff:1977}.

In Table~\ref{tab:values-metals-Lubarda} we present the averages computed according to Lubardas method.

\begin{table}[h!t!b]
{\renewcommand{\arraystretch}{1.1}
\centering
 \begin{tabular}{c|c|c|c|c}
          & Al & Cu & Fe & Nb \\\hline
 $\lambda$[GPa]\!\! & 58.4 & 103.3 & 116.2 & 146.0 \\
 $\mu$[GPa]\!\! & 26.2 & 47.9 & 78.7 & 37.6 \\\hline
 $\ml$[GPa]\!\! & -121 & -278 & -350 & -613 \\
 $\mm$[GPa]\!\! & -273 & -441 & -587 & -206 \\\
 $\mn$[GPa]\!\! & -360 & -630 & -609 & 349
 \end{tabular}
 \caption{We list the averages using Lubardas method~\cite{Lubarda:1997}
 for polycrystals using the single crystal data presented in Table~\ref{tab:values-metals-cubic}.}
 \label{tab:values-metals-Lubarda}
}
\end{table}

\subsection*{A new averaging scheme}

We now address some of the critical points raised above:
In order to be consistent within the averaging scheme of Ref.~\cite{Lubarda:1997}, the present author believes one should replace all strains in the potential \eqref{eq:potential} with $\tilde\cH_{ijkl}\eta_{kl}$ and then vary w.r.t. $\eta_{ij}$ rather than $\eta^0_{ij}$.
This way, the hatted elastic constants will have all their indices contracted with the $\tilde\cH$-tensor and hence will have Voigt symmetry, i.e.
\begin{align}
\tilde C_{ijkl}&= C_{tupq}\tilde\cH_{tuij}\tilde\cH_{pqkl}\,,\nn\\
\tilde C_{ijklmn}&= C_{tupqrs}\tilde\cH_{tuij}\tilde\cH_{pqkl}\tilde\cH_{rsmn}
\,. \label{eq:defHattedCs2}
\end{align}
In the equations above, one should furthermore drop all terms which are higher order in $h$, i.e. all terms proportional to $h^2$ and $h^3$ should be discarded.
In order to reproduce the previous results for the second order constants (assuming the method of Ref.~\cite{Kroener:1958} is correct), one additionally has to use the tensor $\tilde\cH_{ijkl}= \cH_{ijkl}\big|_{h\to h/2}$.
The present averaging will differ from Lubardas (Ref.~\cite{Lubarda:1997}) only for materials with appreciable anisotropy in their SOEC.
Due to the general difficulties mentioned above, one cannot, however, expect the current new averaging scheme to perform much better than the previous ones, and in fact comparing to data it is easily found to be comparable to them, as can be seen from the results presented in Table~\ref{tab:values-metals-myaverages}.

\begin{table}[h!t!b]
{\renewcommand{\arraystretch}{1.1}
\centering
 \begin{tabular}{c|c|c|c|c}
          & Al & Cu & Fe & Nb \\\hline
 $\ml$[GPa]\!\! & -121 & -276 & -355 & -613 \\
 $\mm$[GPa]\!\! & -274 & -461 & -586 & -209 \\
 $\mn$[GPa]\!\! & -359 & -643 & -567 & 346
 \end{tabular}
 \caption{We list the averages for TOEC using the improvements suggested in the present paper for polycrystals using the single crystal data presented in Table~\ref{tab:values-metals-cubic}.}
 \label{tab:values-metals-myaverages}
}
\end{table}

And \eqnref{eq:corrected-strain} exhibits yet another shortcoming of the current averaging schemes for TOEC:
In the corrections for the strain tensor only the anisotropy in the SOEC has been taken into account, but not the anisotropy in the TOEC.
The latter would amount to including additionally a next-to-leading order correction in the strain.
Thus, one might consider replacing $\eta^0_{ij}$ in \eqref{eq:corrected-strain} with $\eta^0_{ij}=\cH_{ijkl}\eta_{kl}+\cH_{ijklmn}\eta_{kl}\eta_{mn}$, as was suggested as a possible future extension of the averaging scheme in Ref.~\cite{Lubarda:1997}.

Another point, which has been raised in the past, see Ref.~\cite{Barsch:1968}, is that even if the polycrystal does not have average rotation elements, the individual grains may well depend on rotational parts of the displacement gradients and this may also explain (part of) the discrepancies between averaged and measured effective TOEC.
If this is true, comparison of averaged with measured effective isotropic elastic constants (using an appropriately generalized averaging scheme) might give us some information about rotation elements in the single crystal grains.

\subsection*{Acknowledgements}

The author thanks D.~J. Luscher and D.~L. Preston for enlightening discussions and D.~J. Luscher for drawing my attention to Ref.~\cite{Lubarda:1997} which became the starting point of this work.
I also thank the anonymous referees for their valuable comments.
This work was performed under the auspices of the U.S. Department of Energy under contract DE-AC52-06NA25396.
In particular, the author is grateful for the support of the Advanced Simulation and Computing, Physics and Engineering Models Program.

\appendix

\setcounter{section}{1}
\setcounter{equation}{0}
\section*{Appendix}

For a crystal of cubic symmetry the SOEC, as defined by the expansion \eqref{eq:stress-strain}, in Cartesian coordinates aligned with the crystal axes read~\cite[p.~435]{Hirth:1982}:
\begin{align}
 C_{ijkl}&=c_{12}\d_{ij}\d_{kl}+c_{44}\left(\d_{ik}\d_{jl}+\d_{il}\d_{jk}\right)-H\d_{ij}\d_{kl}\d_{ik}\,, \nn\\
 H&=2c_{44}+c_{12}-c_{11}
 \,. \label{eq:secondorder-const}
\end{align}
In the isotropic limit $H\to0$, and the three cubic constants reduce to the two Lam{\'e} constants: $c_{12}\to\l$, $c_{44}\to\m$, $c_{11}\to \l+2\m$.
When rotating the SOEC \eqref{eq:secondorder-const} to a different frame, i.e. $C'_{ijkl}=T_{ig}T_{jh}T_{km}T_{ln}C_{ghmn}$ where $T_{ij}$ is a rotation matrix, only the anisotropic part proportional to $H$ changes.


The six TOEC for cubic crystals are typically denoted (in Brugger's notation~\cite{Brugger:1964}) by $c_{111}$, $c_{112}$, $c_{123}$, $c_{144}$, $c_{166}$, and $c_{456}$.
As was the case for the SOEC, their names are inspired by Voigt notation.
Upon introducing three more anisotropy parameters,
\begin{align}
 H_1&\coleq c_{123}+6c_{144}+8c_{456}-3H_2-12H_3-c_{111}\,,\nn\\
 H_2&\coleq c_{123}+2c_{144}-c_{112}\,,\nn\\
 H_3&\coleq c_{144}+2c_{456}-c_{166}
 \,,
\end{align}
the TOEC can be written as
\footnote{
We use a different parametrization than the one given in Ref.~\cite{Lubarda:1997}
--- although both are equivalent ---
which separates more clearly the isotropic from the anisotropic terms,
and which in the isotropic limit coincides with the tensorial expression of~\cite{Toupin:1961}.
}
\begin{align}
  C_{ii'jj'kk'}&=c_{123}\d_{ii'}\d_{jj'}\d_{kk'}\nn\\
 &\quad +c_{144}\Big[\d_{ii'}\left(\d_{jk}\d_{j'k'}+\d_{jk'}\d_{j'k}\right)+\d_{jj'}\left(\d_{ik}\d_{i'k'}+\d_{ik'}\d_{i'k}\right)+\d_{kk'}\left(\d_{ij}\d_{i'j'}+\d_{ij'}\d_{i'j}\right)\Big]\nn\\
 &\quad +c_{456}\Big[\d_{ij}\left(\d_{i'k}\d_{j'k'}+\d_{i'k'}\d_{j'k}\right) +\d_{i'j'}\left(\d_{ik}\d_{jk'}+\d_{ik'}\d_{jk}\right)\nn\\ &\qquad\qquad +\d_{ij'}\left(\d_{i'k}\d_{jk'}+\d_{i'k'}\d_{jk}\right) +\d_{i'j}\left(\d_{ik}\d_{j'k'}+\d_{ik'}\d_{j'k}\right)\Big]
 \nn\\
 &\quad -H_1\d_{ii'}\d_{jj'}\d_{kk'}\d_{ij}\d_{ik}
 -H_2\d_{ii'}\d_{jj'}\d_{kk'}\left(\d_{ij}+\d_{jk}+\d_{ki}\right) \nn\\
 &\quad - H_3\Big[\d_{ii'}\left(\d_{jk}\d_{j'k'}+\d_{jk'}\d_{j'k}\right)\left(\d_{ij}+\d_{ij'}\right)
 +\d_{jj'}\left(\d_{ik}\d_{i'k'}+\d_{ik'}\d_{i'k}\right) \left(\d_{jk}+\d_{jk'}\right)\nn\\
 &\qquad\qquad +\d_{kk'}\left(\d_{ij}\d_{i'j'}+\d_{ij'}\d_{i'j}\right)\left(\d_{ki}+\d_{ki'}\right)\Big]
 \,, \label{eq:cubic-elastic-consts-third}
\end{align}
and $C'_{ii'jj'kk'}=T_{il}T_{i'l'}T_{jm}T_{j'm'}T_{kn}T_{k'n'}C_{ll'mm'nn'}$ in coordinates rotated with respect to the crystal frame, where once again only the anisotropic terms proportional to $H_i$ are affected by $T_{ij}$.
In the isotropic limit, all three anisotropy parameters vanish, $H_i\to0$, implying that
\begin{align}
 c_{166}&=c_{144}+2c_{456}\,, &
 c_{112}&=c_{123}+2c_{144}\,, &
 c_{111}&=
 c_{112} + 4c_{166}
 \,.
\end{align}
Other common parametrizations of the three TOEC in the isotropic limit are the one of Toupin and Bernstein~\cite{Toupin:1961},
\begin{align}
 \n_1&\coleq c_{123}\,, &
 \n_2&\coleq c_{144}\,, &
 \n_3&\coleq c_{456}
 \,,
\end{align}
as well as the one by Murnaghan~\cite{Murnaghan:1937} (see also~\cite{Wallace:1970,Volkov:2015}),
\begin{align}
 \ml&\coleq c_{144}+\tfrac12c_{123} \,, &
 \mm&\coleq 2c_{456}+c_{144} \,, &
 \mn&\coleq 4c_{456}
 \,.
\end{align}
The latter parametrization is the one used in this paper in order to distinguish the (averaged) isotropic constants more clearly from  the cubic ones.

\printbibliography

\end{document}